\documentclass{aa}
\input{psfig}
\usepackage{natbib}
\usepackage{journals}

\def\Real{{\rm I\mathchoice{\kern-0.70mm}{\kern-0.70mm}{\kern-0.65mm}%
  {\kern-0.50mm}R}}

\font \bolditalics = cmmib10
\def\bx#1{\leavevmode\thinspace\hbox{\vrule\vtop{\vbox{\hrule\kern1pt
        \hbox{\vphantom{\tt/}\thinspace{\bf#1}\thinspace}}
      \kern1pt\hrule}\vrule}\thinspace}

\def \vc #1{{\textfont1=\bolditalics \hbox{$\bf#1$}}}

\def\eg{{\bf e}}

\def\thetag{{\vc \theta}}

\def\gammag{{\vc \gamma}}

\def\Cg{{\bf C}}

\def\be{\begin{equation}}
\def\ee{\end{equation}}
\def\ba{\begin{eqnarray}}
\def\ea{\end{eqnarray}}

\def\ltsima{$\; \buildrel < \over \sim \;$}
\def\lsim{\lower.5ex\hbox{\ltsima}}
\def\gtsima{$\; \buildrel > \over \sim \;$}
\def\gsim{\lower.5ex\hbox{\gtsima}}

\begin{document}



   \title{Likelihood Analysis of Cosmic Shear\\ on
 Simulated and VIRMOS-DESCART Data\thanks{Based on 
  observations obtained at the
Canada-France-Hawaii Telescope (CFHT), which is operated by the
National Research Council of Canada (NRCC), the Institut des Sciences
de l'Univers (INSU) of the Centre National de la Recherche
Scientifique (CNRS) and the University of Hawaii (UH), and 
 at the European Southern Observatory telescopes Very Large 
  Telescope (VLT) and the New Technology Telescope (NTT).}}

   \author{ L. Van Waerbeke$^{1,2}$, Y. Mellier$^{1,3}$,
 R. Pell\'o$^4$, U-L. Pen$^{2}$, H.J. McCracken$^{5,6,7}$, B. Jain$^{8}$}
 \offprints{(L. Van Waerbeke) waerbeke@iap.fr}

\institute{$^{1}$Institut d'Astrophysique de Paris, 98 bis boulevard
Arago, 75014 Paris, France,\\
 $^{2}$Canadian Institute for Theoretical Astrophysics, 60 St
George Str., Toronto, M5S 3H8, Canada\\
 $^{3}$Observatoire de Paris, LERMA, 61 Av. de
l'Observatoire, 75014 Paris, France,\\
 $^4$Observatoire Midi-Pyr\'en\'ees, UMR 5572, 14 Avenue E.
                Belin, 31400 Toulouse, France\\
$^5$Department of Astronomy, 
University of Bologna, 
via Ranzani, 1 - 40127 Bologna, Italy\\
$^6$Observatorio Astronomico  di Bologna,
via Ranzani, 1 - 40127 Bologna, Italy\\
 $^7$Laboratoire d'Astrophysique de Marseille, Traverse du Siphon, 13376,
Marseille Cedex 12, France\\
 $^8$Dept of Physics and Astronomy,
University of Pennsylvania,
209 S. 33rd Street, Philadelphia, PA 19104, USA.\\
}
   \markboth{LVW et al}{}


   \markboth{Likelihood Analysis of Cosmic Shear}{}

\authorrunning{LVW et al.}
\abstract{
We present a maximum likelihood analysis of cosmological parameters
from measurements of the aperture mass up to $35$ arcmin, using simulated
and real cosmic shear data. A four-dimensional parameter space is explored
which examines the mean density $\Omega_M$, the mass power spectrum
normalization $\sigma_8$, the shape parameter $\Gamma$ and the redshift of
the sources $z_s$. Constraints on $\Omega_M$ and $\sigma_8$ ({\it resp.}
$\Gamma$ and $z_s$) are then given by marginalizing over $\Gamma$ and
$z_s$ ({\it resp.}  $\Omega_M$ and $\sigma_8$).  For a flat $\Lambda$CDM
cosmologies, using a photometric redshift prior for the sources and $\Gamma
\in [0.1,0.4]$, we find $\sigma_8=\left(0.57\pm0.04\right)
\Omega_M^{\left(0.24\mp 0.18\right) \Omega_M-0.49}$ at the $68\%$
confidence level (the error budget includes statistical noise, full cosmic
variance and residual systematic).  The estimate of $\Gamma$, marginalized
over $\Omega_M \in [0.1,0.4]$, $\sigma_8 \in [0.7,1.3]$ and $z_s$
constrained by photometric redshifts, gives $\Gamma=0.25\pm 0.13$ at
$68\%$ confidence.  Adopting $h=0.7$, a flat universe, $\Gamma=0.2$ and
$\Omega_m=0.3$ we find $\sigma_8$=0.98 $\pm0.06$ . Combined with CMB, our
results suggest a non-zero cosmological constant and provide tight
constraints on $\Omega_M$ and $\sigma_8$. We finaly compare our
results to the
cluster abundance ones, and discuss the possible discrepancy with
the latest determinations of the cluster method. In particular we point
out the actual limitations of the mass power spectrum prediction in the
non-linear regime, and the importance for its improvement.
}

\maketitle

\section{Introduction}

In the standard cosmological picture, the structures in the Universe
grow from the gravitational collapse of initial Gaussian density
perturbations. The properties of mass distribution at low redshift are
expected to express the latest and one of the most explicit footprint
of the formation process, so their description from cosmological
surveys can be challenged against theoretical predictions resulting
from this paradigm. For example, a direct observation of the mass
distribution in structures is believed to be an unequivocal test of
cosmological scenario of structure formation.  If so, the weak
gravitational lensing produced on distant galaxies by large scale
structures is a direct probe of dark matter, regardless the light
distribution. It is therefore a robust technique to challenge the current
cosmological models.  In particular, it can reliably probe small
angular scale and look into the transition to the quasi-linear and
non-linear regimes, where comparison between observations and
cosmological models are still difficult.

The cosmological origin of the coherent distortion fields detected in
cosmic shear surveys is now firmly established
\citep{2000MNRAS.318..625B,2001astro.ph.10210H,2002astro.ph..2285H,kaiser2000,
  2001A&A...368..766M,2001astro.ph..9182P,2001ApJ...552L..85R,
  2000A&A...358...30V,2001A&A...374..757V,2000Natur.405..143W}.
\cite{2001A&A...374..757V} have shown that the measurements provided by
different statistical estimators of distortion signal are consistent
with the gravitational lensing hypothesis with a high confidence level,
so that present-day data can already constrain cosmological parameters.
Their joint estimate of the mass density $\Omega_M$ and the power
spectrum normalization $\sigma_8$ led to consistent results with the
cluster abundance constraints \citep{2001MNRAS.325...77P} and confirmed
earlier tentative obtained by \cite{2001A&A...368..766M} and
\cite{2001ApJ...552L..85R} using ESO-VLT/CFHT and HST data
respectively. A recent measurement done on a shallow survey (therefore
very different in depth) confirmed also this agreement
\citep{2001astro.ph..9514H,2002astro.ph..2285H}.

So far, the cosmological parameter estimation from cosmic shear relied
on prior knowledge of the slope of the mass power spectrum $\Gamma$
and/or the mean redshift $\bar z_s$ of the lensed galaxy population. In
fact, the statistical properties of cosmic shear depends significantly
on these quantities \citep{1992ApJ...388..272K,1997A&A...322....1B,
  1997ApJ...484..560J}, so any prior on these parameters may have a
serious impact on the cosmological parameter estimation.  For instance
changing the shape of the power spectrum in either direction would
favor low or high matter densities, by changing the normalization
accordingly.  This ambiguity expresses a degeneracy between the
normalization and the mass density, which depends on the choice of
$\Gamma$ \citep{2001A&A...374..757V}.  \cite{1997ApJ...484..560J}
addressed this issue by pointing out that a measurement of the cosmic
shear in both linear {\it and} non-linear scales could break the
degeneracy, so that one in principle recover simultaneously $\Gamma$,
$\sigma_8$ and $\Omega_M$ from the shear variance alone. Unfortunately,
the redshift of the sources is also a strongly degenerate parameter
with $\sigma_8$, which definitely hampers shear variance analysis to
provide unequivocal discrimination of cosmological models.  In fact,
stringent constraints on the cosmological parameters from the shear
variance are possible only with an accurate knowledge of the source
redshifts and a measurement which extends over a large range of scales.

In this paper we carry out a full maximum likelihood analysis of cosmic
shear data over the four parameters $\Omega_M$, $\sigma_8$, $\Gamma$,
and $\bar z_s$ for flat and open cosmologies. Using both simulations
and observations, we study slices and projections in this parameter
space and discuss the reliability of cosmological constraints which are
based on catalogues having similar size and depth as current cosmic
shear surveys. In particular we give an estimate of $\Omega_M$ and
$\sigma_8$ by marginalizing over the power spectrum shape and sources
redshifts.  The improvement of our knowledge of the source redshift is
crucial to gain better accuracy for the other parameters.  In this work
we use photometric redshifts \footnote{derived from other data sets} to
put priors on the source redshift distribution.

The paper is organized as follows. Section 2 presents a brief summary
of some theoretical concepts and introduces the shear estimators used
throughout the paper.  Section 3 describes the data and how shear
quantities were obtained from the survey catalogue.  The likelihood
method and the details of the priors are presented in Section 4.
Section 5 shows and discusses the results on the parameter estimates on
both simulated and real surveys. Finally, conclusions are presented in
Section 6.

\section{Theory}

Following the notation in \cite{1998MNRAS.296..873S}, we define the
power spectrum of the convergence as

\begin{eqnarray}
P_\kappa(k)&=&{9\over 4}\Omega_0^2\int_0^{w_H} {{\rm d}w \over a^2(w)}
P_{3D}\left({k\over f_K(w)};
w\right)\times\nonumber\\
&&\left[ \int_w^{w_H}{\rm d} w' n(w') {f_K(w'-w)\over f_K(w')}\right]^2,
\label{pofkappa}
\end{eqnarray}
where $f_K(w)$ is the comoving angular diameter distance out to a
distance $w$ ($w_H$ is the horizon distance), and $n(w(z))$ is the
redshift distribution of the sources given in Eq.(\ref{zsource}).
$P_{3D}(k)$ is the non-linear mass power spectrum computed according to
\cite{1996MNRAS.280L..19P}, and $k$ is the 2-dimensional wave vector
perpendicular to the line-of-sight.  The top-hat shear variance
(smoothing window of radius $\theta_c$) and the shear correlation
function can be written as

\begin{equation}
\langle\gamma^2\rangle={2\over \pi\theta_c^2} \int_0^\infty~{{\rm d}k\over k} P_\kappa(k)
[J_1(k\theta_c)]^2,
\label{theovariance}
\end{equation}

\begin{equation}
\langle\gamma\gamma\rangle_\theta={1\over 2\pi} \int_0^\infty~{\rm d} k~
 k P_\kappa(k) J_0(k\theta).
\label{theogg}
\end{equation}
Because the weak distortion field can be generated by non-lensing
mechanisms, it is important to measure separately the $E$ and $B$
components of the shear.  These modes were introduced initially to test
for the gravitational origin of the lensing signal
\citep{2000astro.ph.12336C} since a potential gravitational field is
expected to produce only curl-free shear patterns ($E$ mode). Any
measurable $B$ mode should be interpreted as a measurement of residual
systematic in the data (Point Spread Function correction, intrinsic
alignment or anything else) and must be removed from the weak lensing
signal prior to cosmological interpretation of cosmic shear data.

The extraction of both modes is not trivial. The $E$ and $B$-modes
decompositions of the top-hat shear variance, and of the shear
correlation function given in Eq.(\ref{theovariance},\ref{theogg}) are
only defined up to a integration constant
\citep[see][]{2000astro.ph.12336C, 2001astro.ph..9182P}. This constant
depends on the extrapolated cosmic shear signal either at small ($<30$
arc-seconds) or large ($>1$ degree) scales.  These boundary conditions
turn out to be a severe limitation which hampers reliable derivations
of both modes from our present-day data because we do not cover yet
very large angular scales and we still suffer from systematics on very
small scales which are not well understood.  As pointed out by
\cite{2001astro.ph..9182P}, the only unambiguous $E$ and $B$ mode
decomposition is carried out by the aperture mass, $M_{ap}$:

\begin{equation}
M_{\rm ap}=\int_{\theta < \theta_c}~{\rm d}^2\thetag \kappa(\thetag)~U(\theta),
\end{equation}
where $\kappa(\thetag)$ is the convergence field, and $U(\theta)$ is the
zero mass aperture window \citep{1998MNRAS.296..873S}:

\begin{equation}
U(\theta)={9\over \pi \theta_c^2} \left(1-{\theta^2\over\theta_c^2}\right)
\left({1\over 3}-{\theta^2\over\theta_c^2}\right).
\label{Ufilter}
\end{equation}
This estimator was introduced in \cite{kaiser94} to study clusters of
galaxies, but it also has an important potential for cosmic shear
analysis \citep{1998MNRAS.296..873S}.

$\langle M_{\rm ap}^2\rangle$ can be calculated directly from the shear
$\gammag$ without the need for a mass reconstruction. It can be
rewritten as a function of the shear if we express
$\gammag=(\gamma_t,\gamma_r)$ in the local frame of the line connecting
the aperture center to the galaxy. $M_{\rm ap}$ can therefore be
expressed as function of $\gamma_t$ only
\citep{1991ApJ...380....1M,1992ApJ...388..272K}:

\begin{equation}
M_{\rm ap}=\int_{\theta < \theta_c}~{\rm d}^2\thetag \gamma_t(\thetag)~Q(\theta),
\label{mapfromshear}
\end{equation}
where the filter $Q(\theta)$ is given from $U(\theta)$:

\begin{equation}
Q(\theta)={2\over \theta^2}\int_0^\theta~{\rm d}\theta'~\theta'~U(\theta')-U(\theta)
\label{Qfct}
\end{equation}
The aperture mass variance is related to
the convergence power spectrum (Eq.\ref{pofkappa})by:

\begin{equation}
\langle M_{\rm ap}^2\rangle={288\over \pi\theta_c^4} \int_0^\infty~{{\rm d}k\over k^3}
 P_\kappa(k) [J_4(k\theta_c)]^2.
\label{theomap}
\end{equation}
The $B$-mode is obtained by replacing $\gamma_t$ with $\gamma_r$ in
Eq.(\ref{mapfromshear}).  Although this estimator is robust and does
not depend on an unknown integration constant, it is less sensitive to
cosmological parameters than the top-hat variance or the shear
correlation functions \citep{2001A&A...374..757V}.

\section{Measurements}

We use the observations done within the VIRMOS-DESCART project
\footnote{http://terapix.iap.fr/DESCART} by the VIRMOS
\footnote{http://www.astrsp-mrs.fr} imaging and spectroscopic survey.
The data cover an effective area of 8.5 sq.deg. in the I-band, with a
limiting magnitude $m_{I_{AB}}=24.5$. Technical details of the data set
are given in \cite{2001A&A...374..757V}. We applied a bright magnitude
cut at $m_I=21$ in order to exclude the foreground objects from the
source galaxies.  The shape of the galaxies are measured and analyzed
as described in this paper, to which we refer for technical details.

The location of the $i$-th galaxy is given by $\thetag_i$, the
ellipticity by $\eg(\thetag_i)=(e_1,e_2)$, and its weight $w_i$.  The
ellipticity is an unbiased estimate of the shear $\gammag(\thetag_i)$.
The quantity measured on the data are the binned tangential and radial
shear correlation functions.  They are given by a sum over galaxy pairs
$(\thetag_i,\thetag_j)$

\begin{eqnarray}
\xi_{tt}(r)&=&{\displaystyle\sum_{i,j} w_i w_j e_t(\thetag_i)\cdot e_t(\thetag_j)
\over \displaystyle\sum_{i,j} w_i w_j}\nonumber \\
\xi_{rr}(r)&=&{\displaystyle\sum_{i,j} w_i w_j e_r(\thetag_i)\cdot e_r(\thetag_j)
\over \displaystyle\sum_{i,j} w_i w_j},
\label{corrfct}
\end{eqnarray}
\begin{figure}
\centerline{
\psfig{figure=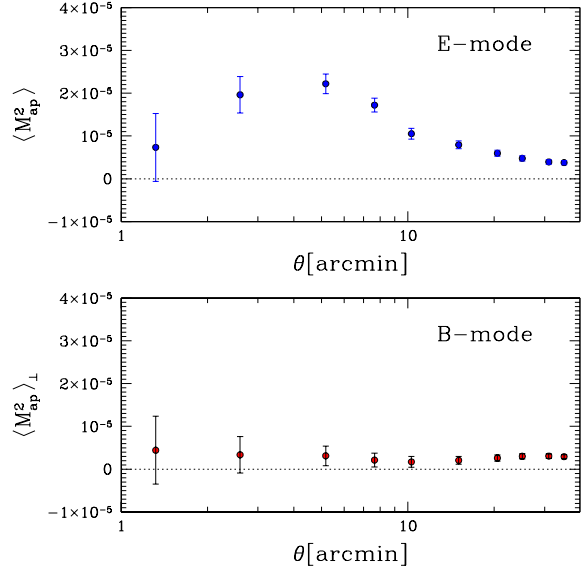,height=8cm}}
\caption{\label{map_signal.ps} Above is the aperture mass statistic
$\langle M_{\rm ap}^2\rangle$ ($E$-mode) and below the aperture mass
$\langle M_{\rm ap}^2\rangle_\perp$ computed with galaxies rotated by $45$ degrees
($B$-mode). Error bars are $1-\sigma$ statistical errors.
}
\end{figure}
where $r=|\thetag_i-\thetag_j|$, and $(e_t, e_r)$ are the tangential and radial
ellipticities defined in the frame of the line connecting a pair of galaxies.

From Eq.(\ref{corrfct}), we define $\xi_+(r)$
and $\xi_-(r)$ which are respectively the sum and the difference of the two
correlation functions:

\begin{equation}
\xi_+(r)=\xi_{tt}(r)+\xi_{rr}(r); \ \
\xi_-(r)=\xi_{tt}(r)-\xi_{rr}(r).
\label{xipm}
\end{equation}
Both $\xi_+(r)$ and 
$\xi_-(r)$ are computed
from a summation of the correlation function defined in Eq.(\ref{xipm}), while
 the $E$ and $B$ modes  aperture mass are derived by integration of the
correlation functions with an appropriate window 
 (see  \cite{2000astro.ph.12336C} for  
  general derivations and \cite{2001astro.ph..9182P} for a 
 practical application to our filter).

The $E$ mode aperture mass is

\begin{equation}
\langle M_{\rm ap}^2\rangle=\pi\int_0^{2\theta_c} rdr {\cal W}(r) \xi_+(r)+
\pi\int_0^{2\theta_c} rdr \tilde{\cal W}(r) \xi_-(r),
\label{MapE}
\end{equation}
where ${\cal W}(r)$ and $\tilde{\cal W}(r)$ are given in \cite{2000astro.ph.12336C}\footnote{
Useful expressions using similar formalism as this work 
   can be found in \cite{2001astro.ph.12441S}}. The $B$-mode
is obtained by changing the sign of the second term in Eq.(\ref{MapE}) (which is
equivalent to the $45$ degrees rotation test, or else $\gamma_t\rightarrow\gamma_r$).

Figure \ref{map_signal.ps} shows the $E$ mode (top) and $B$ mode
(bottom) measured in our galaxy sample. Using the $B$-mode measurement,
we found the source of the residual systematics at $3-4'$ reported by
\cite{2001A&A...374..757V} and \cite{2001astro.ph..9182P}: it was
caused by the third order polynomial fit to the PSF, which produced
wings at the edge of the CCDs.  A second order fitting removed most of
the unwanted $B$ mode contribution without spoiling the $E$ mode
signal.  As shown in Figure \ref{map_signal.ps}, the residual
systematics are consistent with zero up to 10 arc-minutes and remains
flat over the whole angular scale explored by the data.  Clearly, the
signal is dominated by the $E$ mode contribution at least up to 25
arc-minutes.  This demonstrates that signal produced by intrinsic
alignment of galaxies is not detected at this level.
 
\section{Parameter Estimation}

\subsection{Redshift distribution of galaxies in the VIRMOS-DESCART data}

We estimated the redshift distribution of our catalogue from a
combination of the Hubble Deep Fields North and South data
\citep{1999ApJ...513...34F,1998astro.ph..12339C} and VLT observations
of the cluster MS1008-1224. Both HDF and MS1008-1224 observations are
much deeper than the cosmic shear sample of galaxies considered in this
work, so that magnitude measurements and photometric redshift estimates
up to $I_{AB}=24.5$ are based on reliable data with high
signal-to-noise ratio.

The VLT MS1008-1224 galaxy sample comprises deep $UBVRI$ observations,
carried out by the Science Verification Team (SVT) at ESO/VLT with the
FORS1 and FORS2 instruments \citep{1998Msngr..94....1A} and deep $J$
and $K$ data obtained at the ESO/NTT with SOFI (Program 66.A-0316(A);
PI Mellier). The extension of early deep SVT observations to U band
with FORS2 and more recently to near infrared with SOFI, which has
similar field of view as FORS (5.5 arc-minutes against 6.8
arc-minutes), allow us to considerably improve the accuracy of
photometric-redshifts of foreground, cluster and background galaxies
over the whole field.  As compared to HDF the VLT/NTT observations are
not as deep, but they provide a much larger sample of galaxies because
they cover of field of view 15 times larger than HDF.  In total, 920
galaxies having $I_{AB} \le 24.5$ and $UBVRIJK$ data have been added to
HDF data.

The deep $UBVRI$ data are described at the ESO
site\footnote{http://www.hq.eso.org/science/ut1sv } and in
\cite{Athreyaetal02}. A complete description of the new $J$ and $K$
band data will be presented elsewhere \citep{gavazzi}. In brief, the
exposure times of NTT/SOFI $J$ and $K$ bands were 5h30 and 6h
respectively. The completeness limits are $J=23.$ and $K=22.$ and both
$J$ and $K$ complete samples encompasses more than 90\% of the $I_{AB}
\le 24.5$ galaxies. Hence, most galaxies used for determining the
photo-$z$ distribution of galaxies up to $I_{AB} \le 24.5$ have
reliable $J$ and $K$ photometric measurements to secure a collection of
redshift on a very large sample of galaxies, which covers a broad
magnitude range homogeneously spread over the whole field.  The
presence of the lensing cluster ($z=0.306$) in the field only affects
the redshift range $0.26<z<0.36$. These data have been removed from the
sample and the redshift distribution interpolated in this redshift
range.  The magnification bias may also change the redshift
distribution of galaxies inside the very center of the cluster where
the gravitational depletion is significant
\citep[see][]{Athreyaetal02}.  We therefore also removed the central
part ($R<40$ arc-second) of the cluster from the sample. Since this
region is also the most contaminated by the brightest cluster members,
the depletion itself turns out to have no impact of the galaxy
selection criterion.

The photometric redshifts ($z_{phot}$) were measured using the fitting
algorithm $hyperz$ developed by \cite{Bolzonellaetal00}.  Each
$z_{phot}$ is inferred by comparing the spectral energy distribution of
galaxies, as sampled by their $UBVRIJK$ photometric flux, to a set of
spectral templates representative of common late and early type
galaxies which are followed with look-back time according to Bruzual \&
Charlot's evolution models \citep[GISSEL98;][]{BruzualCharlot93}.  The
validation of $hyperz$ is discussed at length in
\cite{Bolzonellaetal00}. It has been conclusively gauged against
spectroscopic redshift on MS1008-1224 data by \cite{Athreyaetal02}.
Details on photometric redshift techniques can be found in those
papers.

\begin{figure}
\centerline{
\psfig{figure=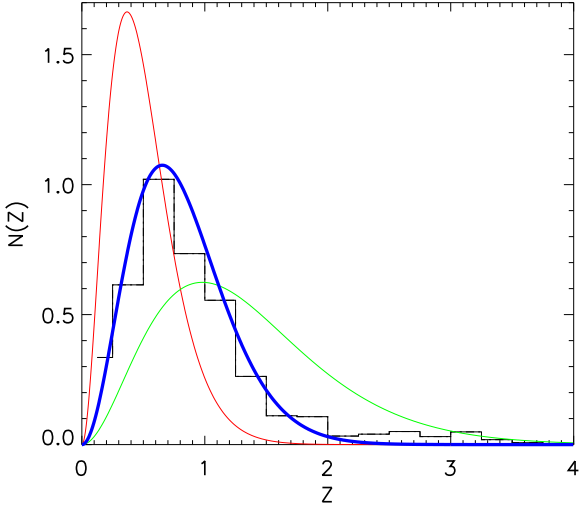,height=7cm}}
\caption{\label{allndez.ps} The histogram shows the photometric redshift
distribution from the joint VLT and HDF fields.
The thick solid line is the theoretical $n(z)$ from
Eq.(\ref{zsource}) with $z_s=0.44$. The low and high redshift thin solid
lines correspond to $z_s=0.24$ and $z_s=0.64$, our extreme redshift
distributions used in this paper.
}
\end{figure}

The compiled photometric distribution is shown on Fig.\ref{allndez.ps}.
For the purpose of marginalization we parameterize this distribution
with the following normalized function:

\begin{equation}
n(z)={\beta\over z_s \ \Gamma\left({1+\alpha\over \beta}\right)} 
\left({z\over
z_s}\right)^\alpha \exp\left[-\left({z\over z_s}\right)^\beta\right],
\label{zsource}
\end{equation}
where $\alpha=2$ and $\beta=1.2$. For these values of $\alpha$ and
$\beta$, the mean redshift is $\bar z_s \approx 2.1\ z_s$ and the
median redshift is $\approx 1.9\ z_s$. We allowed $z_s$ to vary from
$0.24$ to $0.64$, which corresponds to a mean redshift varying from
$0.5$ to $1.32$. These two extreme models are shown on
Fig.\ref{allndez.ps}: they are clearly conservative bounds on the
redshift distribution in the data. The curve on Fig.\ref{allndez.ps}
shows the best fit model, with $z_s=0.44$ ($\bar z_s = 0.9$).

\subsection{Maximum Likelihood}

The dominant cosmological parameters for the 2-point cosmic shear
statistics are the mean mass density $\Omega_M$, the power spectrum
normalization $\sigma_8$, the shape parameter $\Gamma$ and the redshift
of the sources \citep[see][]{1997A&A...322....1B,
  1997ApJ...484..560J,1999A&A...342...15V} Our parameter space has
therefore four dimensions, but we truncate the exploration volume
inside a realistic range defined as $\Omega_M \in [0.1,1]$, $\sigma_8
\in [0.3,1.6]$ and $\Gamma \in [0.05,0.7]$ with a sampling of $10\times
14\times 14$.  For the analysis of the VIRMOS-DESCART data, the
redshift of sources is parameterized by Eq.(\ref{zsource}) with $z_s \in
[0.24,0.64]$ and a sampling of $9$.  For the simulations, the sources
are placed at redshift unity, therefore in the maximum likelihood
analysis we assumed we knew the shape (Dirac distribution), but we
allowed the redshift $z_s$ to vary between $0.7$ and $1.2$ (sampling of
$6$).  In fact we found that the real shape of the source distribution
does not matter, but the agreement with the mean redshift does.  This
parameter range box ($\Omega_M$, $\Gamma$, $\sigma_8$, $z_s$) defines
what we call the {\it default prior box}.  The model predictions are
then interpolated with an oversampling seven times better in each
dimension.

\begin{figure}
\centerline{
\psfig{figure=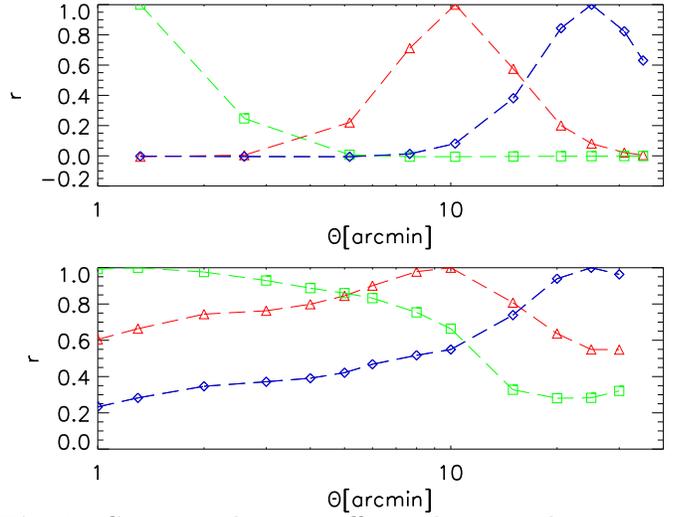,height=7cm}}
\caption{\label{crosscorrel.ps} Cross-correlation coefficient between three
measurement scales for the statistical noise (above) and the cosmic variance
(below). The three scales are $1.3'$ (squares), $10.3'$ (triangles) and
$25.1'$ (diamonds). The cosmic variance cross-correlation is obtained from
ray-tracing simulations (\cite{2000ApJ...530..547J}).
}
\end{figure}

Let $d_i$ be the data vector ({\it i.e.} the aperture mass $\langle
M_{\rm ap}^2\rangle$ for different scales $\theta_i$), and
$m_i(\Omega_M,\sigma_8,\Gamma,z_s)$ the model predictions.  The
likelihood function of the data is

\begin{equation}
{\cal L}={1\over (2\pi)^n|\Cg|^{1/2}} \exp\left[(d_i-m_i)\Cg^{-1}(d_i-m_i)^T\right],
\end{equation}
where $n=10$ is the number of scales and $\Cg$ is the $10\times 10$
covariance matrix,

\begin{equation}
C_{ij}=\langle (d_i-m_i)^T(d_j-m_j)\rangle.
\end{equation}
$\Cg$ can be decomposed as $\Cg=\Cg_n+\Cg_s+\Cg_b$, where $\Cg_n$ is the
statistical noise, $\Cg_s$ the cosmic variance covariance matrix
and $\Cg_b$ the residual
bias. $\Cg_n$ has been measured in \cite{2001astro.ph..9182P}, so  we
just reproduce here its general behavior: the top panel on
figure \ref{crosscorrel.ps} shows the cross-correlation coefficient for
$2$, $10$, and $35$ arc-min with the other scales.
In order to account for residual systematics, we decided to add
quadratically the residual $B$ mode (see the bottom panel in Figure
\ref{map_signal.ps}) to the error of the signal. Given that there is no
clearly identified scheme to deal with the residual systematics yet,
this appears to be the safest and most conservative attitude. The
diagonal part of the bias correlation matrix $\Cg_b$ is therefore given
by the $B$ mode signal, and the off-diagonal terms follow the same
correlation properties as the $E$ mode (the $E$ and $B$ covariance
matrices for the statistical noise are actually identical). 
\\
The cosmic variance
covariance matrix $\Cg_s$ is trickier to estimate. Assuming the field
is Gaussian is too simplistic, since the observed scales are within the
non-linear and weakly non-linear regimes, so in principle a complete
description of non-Gaussian contributions to the error budget cannot be
carried out without detailed cosmological simulations.  In order to
avoid this heavy procedure, we focused instead on a simpler alternative
based on non-linear perturbation theory. It was pointed out in
\cite{1999ApJ...527....1S} that, for the convergence power spectrum,
the ratio of the Gaussian to Non-Gaussian errors is almost independent
of scale, and close to $1$ for any cosmology.  We investigated whether
this statement could be also valid in real space, using three
ray-tracing simulations for three different cosmological models
\citep{2000ApJ...530..547J}.
\begin{figure}
\centerline{
\psfig{figure=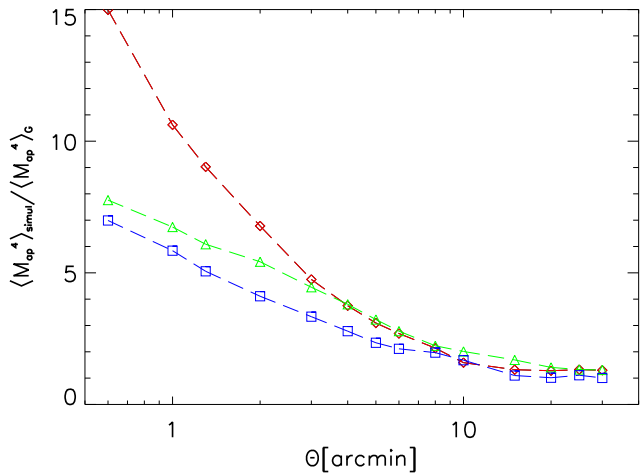,height=7cm}}
\caption{\label{cosvar.ps} Ratio between the aperture mass dispersion obtained from
ray-tracing simulations (\cite{2000ApJ...530..547J}) and the dispersion obtained
from the Gaussian field hypothesis for a survey of similar size. The three curves
correspond to a $\tau$CDM (squares), OCDM (triangles) and $\Lambda$CDM (diamonds),
showing that the ratio is little dependent of the cosmological model above $3'$.
paper.}
\end{figure}
Figure \ref{cosvar.ps} shows this ratio for $\Lambda$CDM, $\tau$CDM and
OCDM. For scales larger than $3$ arcmin., it is indeed nearly
independent of cosmology.  At smaller scales the $\Lambda$CDM model has
a larger cosmic variance, but this is not important, since below a few
arcminutes statistical noise dominates (see Figure
\ref{data_models.ps}). Therefore, although the result in
\cite{1999ApJ...527....1S} is clearly not valid for very small scales,
it is still weakly sensitive to cosmology. We then approximated $\Cg_s$
in the following way: we compute the Gaussian cosmic variance for each
model, then we convert it to a non-Gaussian cosmic variance using the
average kurtosis in Figure \ref{cosvar.ps}. The cross-correlation
coefficient is taken from the ray-tracing simulations. The different
scales are rather correlated, as shown on Figure \ref{crosscorrel.ps}
(bottom panel).  As we shall see in Figure \ref{data_models.ps}, even a
wrong estimate of the cosmic variance by a factor of two has no
consequences on our parameter estimate, given that the errors are
dominated by $\Cg_n$ and $\Cg_b$.

\section{Applications}

We now apply the likelihood analysis to simulated sky images and to the
VIRMOS-DESCART data.

\subsection{Mock catalogues}

The mock catalogues are generated from simulated sky images
following the procedure described in \citep{2001A&A...366..717E}
a simulated catalogue of galaxies is first lensed and then 
used to generate a CCD image of the sky. But 
instead of having a constant shear amplitude on each 
field,  the distortion of galaxies is introduced 
using ray-tracing simulations \citep{2000ApJ...530..547J}

As for real sky surveys, the mock catalogues contain the following
features
\begin{itemize}
\item galaxy intrinsic shape fluctuations; \item masks; \item
noise from galaxy shape measurements and systematics from PSF
corrections \ ,
\end{itemize}
\begin{figure}
\centerline{
\psfig{figure=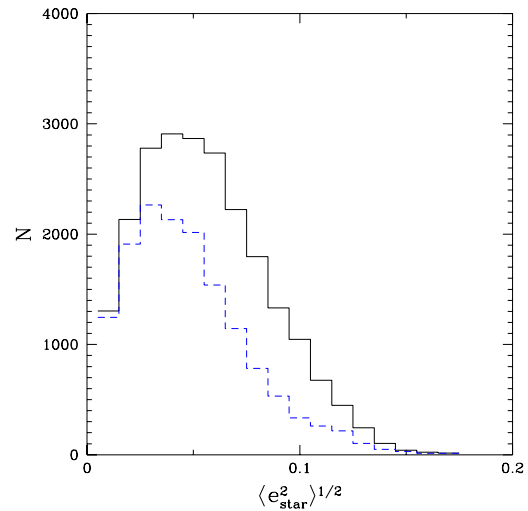,height=7cm}}
\caption{Histogram of the Point Spread Function anisotropy of the
stars in the simulated images (solid line) and in the
VIRMOS-DESCART survey (dashed line).}
\label{psf_aniso}
\end{figure}
and the simulated images reproduce similar observational conditions
 as real data (PSF anisotropy, limiting magnitude,
luminosity functions, galaxy and star number densities, intrinsic
ellipticity...).  The  simulated galaxies are then analyzed exactly in the same
way as real data, following the procedure described in
\citep[ 2001b]{2000A&A...358...30V}

\begin{figure*}
\centerline{
\psfig{figure=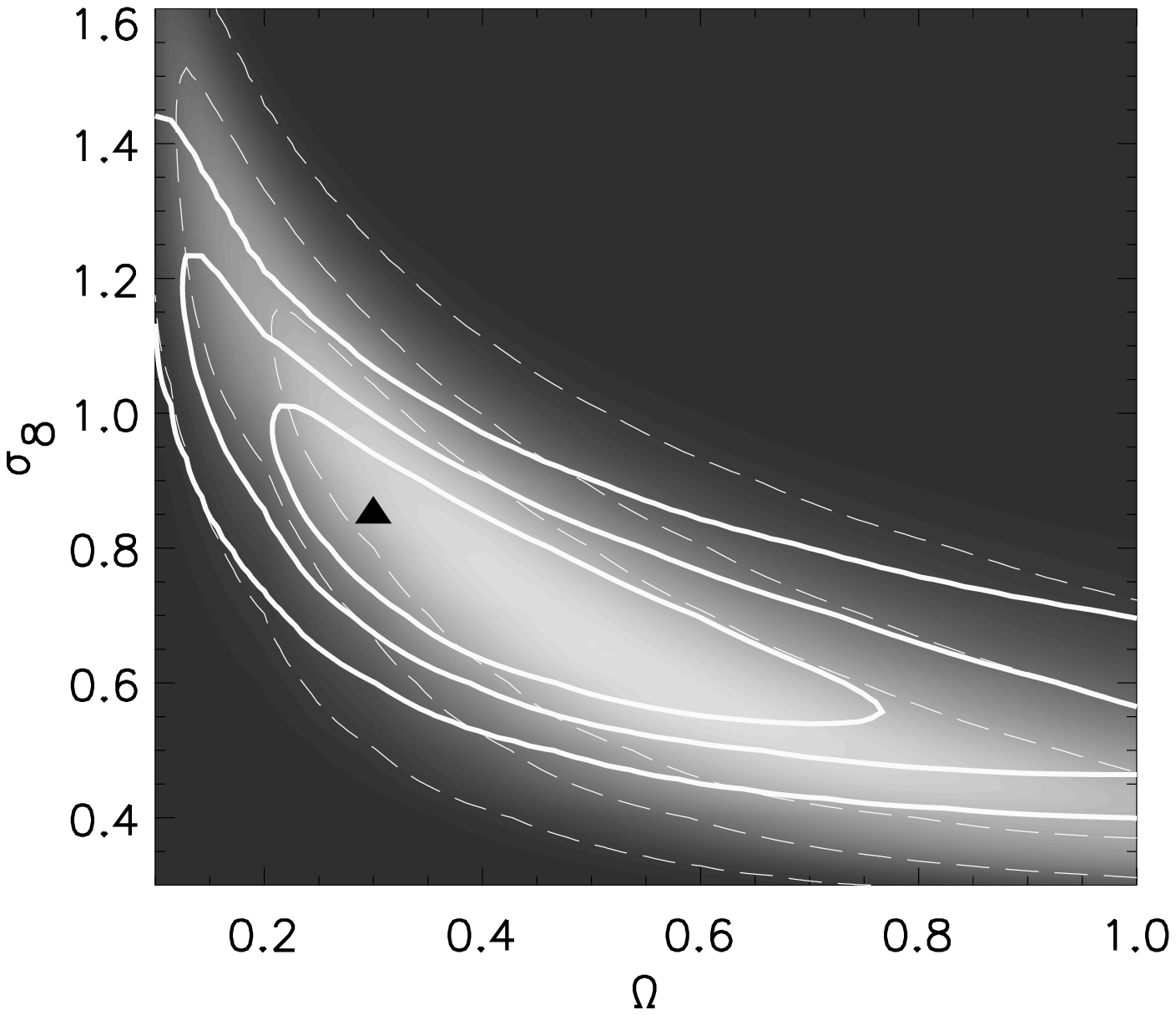,height=7cm}
\psfig{figure=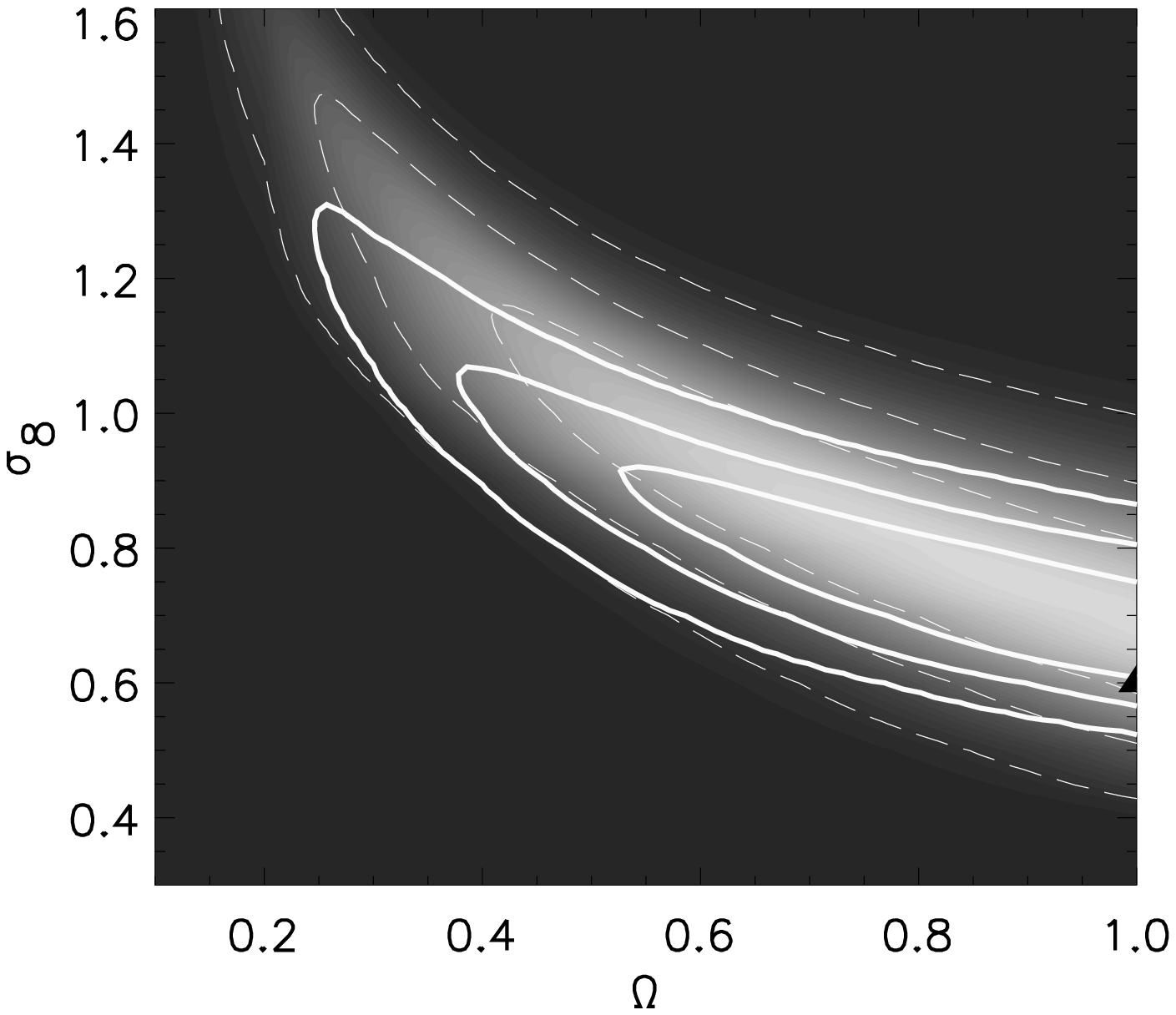,height=7cm}
}
\caption{\label{tcdm_flat.ps} Constraints on $\sigma_8$ and $\Omega_M$ for the
OCDM (left) and $\tau$CDM (right) simulations. The gray levels and the dashed
contours show the -simulation- default prior constraints, with confidence
levels of $0.68$,
$0.95$ and $0.999$. The true model, indicated by a dark triangle, is
$\Omega=0.3$, $\sigma_8=0.85$ (left) and $\Omega=1$, $\sigma_8=0.6$ (right). The
thick solid line contours are for a prior $\Gamma\in[0.1,0.3]$ and
$\bar z_s \in [0.9,1.1]$.
}
\end{figure*}

\begin{figure}
\psfig{figure=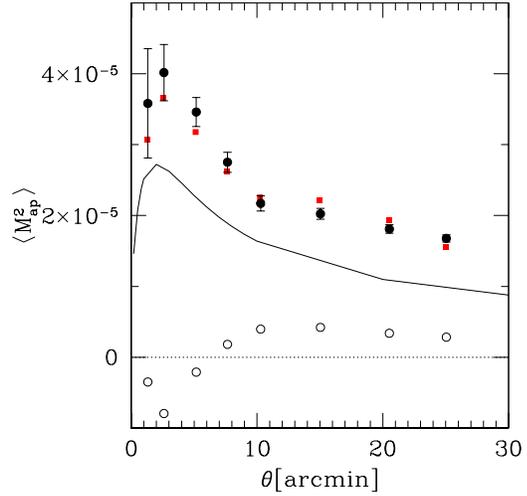,height=7cm}
\caption{\label{fit_model.ps} The filled circles with error bars show the
aperture mass measured on the $\tau$CDM simulated sky images,
while the filled small squares show the signal measured in
the input catalogue. The open circles
show the measured residual $B$-mode. Measurements and simulation
are in perfect agreement, but the non-linear prediction obtained from
\cite{1996MNRAS.280L..19P} for this model
(solid line) is significantly off.
}
\end{figure}

We used two ray-tracing simulations from \cite{2000ApJ...530..547J}:
one is OCDM, as described in Section 3, and the other is a $\tau$CDM
with $\Gamma=0.21$ and $\Omega=1$.  For each simulation we produced
$11$ square degrees of simulated sky images containing roughly $30$
galaxies per arcmin$^2$, with a pixel size of $0.2$ arcsec.  Figure
\ref{psf_aniso} compare the star anisotropy between the simulated
fields (solid line) and the data (dashed line). The likelihood function
is computed for 11760 models ($10\times 14 \times 14 \times 6$).
Figure \ref{tcdm_flat.ps} shows the results for the maximum likelihood
analysis of these two simulation sets. We clearly converge to the right
cosmological model, which validates our likelihood approach for the
data (see Section 5.2). However, we should point out here that the
likelihood method assumes that our theoretical predictions are accurate
compared to the precision of the measurements. Our simulations shows
this is unfortunately not necessarily the case with today's lensing
data sets. For instance, it was shown in \cite{2001MNRAS.322..918V}
(Figure 2), that the non-linear predictions fails badly for the
aperture mass with a $\tau$CDM model.  This failure should not be a
surprise: it was already noticed in the projected power spectrum in
\cite{2000ApJ...530..547J} (Figure 8), and even the VIRGO simulations
\citep[see][ Figure 7]{1998ApJ...499...20J} already noticed a
mismatch between the 3D non-linear predictions and the measured power
spectrum. In the case of our $\tau$CDM simulation, the potential
problem is an overestimate of the power spectrum normalization
$\sigma_8$.  This is illustrated in Figure \ref{fit_model.ps} where we
compare the measured power to the Peacock \& Dodds prediction for that
model. With a smaller statistical error and/or residual bias and cosmic
variance, the true model with $\sigma_8=0.6$ will become excluded from
our $3-\sigma$ contours in the right panel of figure
\ref{tcdm_flat.ps}. In fact a maximum likelihood analysis on the noise
free catalogue would give $\sigma_8=0.8$, that is $20\%$ larger than
the true $\sigma_8$, which corresponds to the lack of power in the
predicted non-linear signal. We will get back to a more detailed
discussion in Section 6 about this problem.

\subsection{VIRMOS-DESCART data}
We first consider flat cosmologies, since this is the class of models
currently favored by the cosmic microwave background measurements
\citep{2000Natur.404..955D}, but alternative open universes are also
investigated. In either case, the likelihood function is computed for
17640 ($10\times 14\times 14 \times 9$) models using
Eq.(\ref{theomap}), as a function of angular scale, and for a regular
spacing in the default prior box.

Figure \ref{flat.ps} shows the four parameters constraints for
different priors and marginalized parameters for the flat cosmology.
The dashed lines shows the $68\%$, $95\%$ and $99.9\%$ contours when
the default prior is applied for the two remaining parameters. We
cannot extract strong constraints in that case, but the right panel
shows an interesting tendency between $\Gamma$ and $z_s$: a flat power
spectrum (large $\Gamma$) can account for an underestimated source
redshift.  The thick solid curves shows the same contours with a
stronger prior: $\Gamma$ and $z_s$ are marginalized over $[0.1,0.4]$
and $[0.39,0.54]$ for the left panel, and $\Omega_M$ and $\sigma_8$ are
marginalized over $[0.1,0.4]$ and $[0.7,1.3]$ for the right panel. We
obtain the following constraint from the left panel:

\begin{figure*}
\centerline{
\psfig{figure=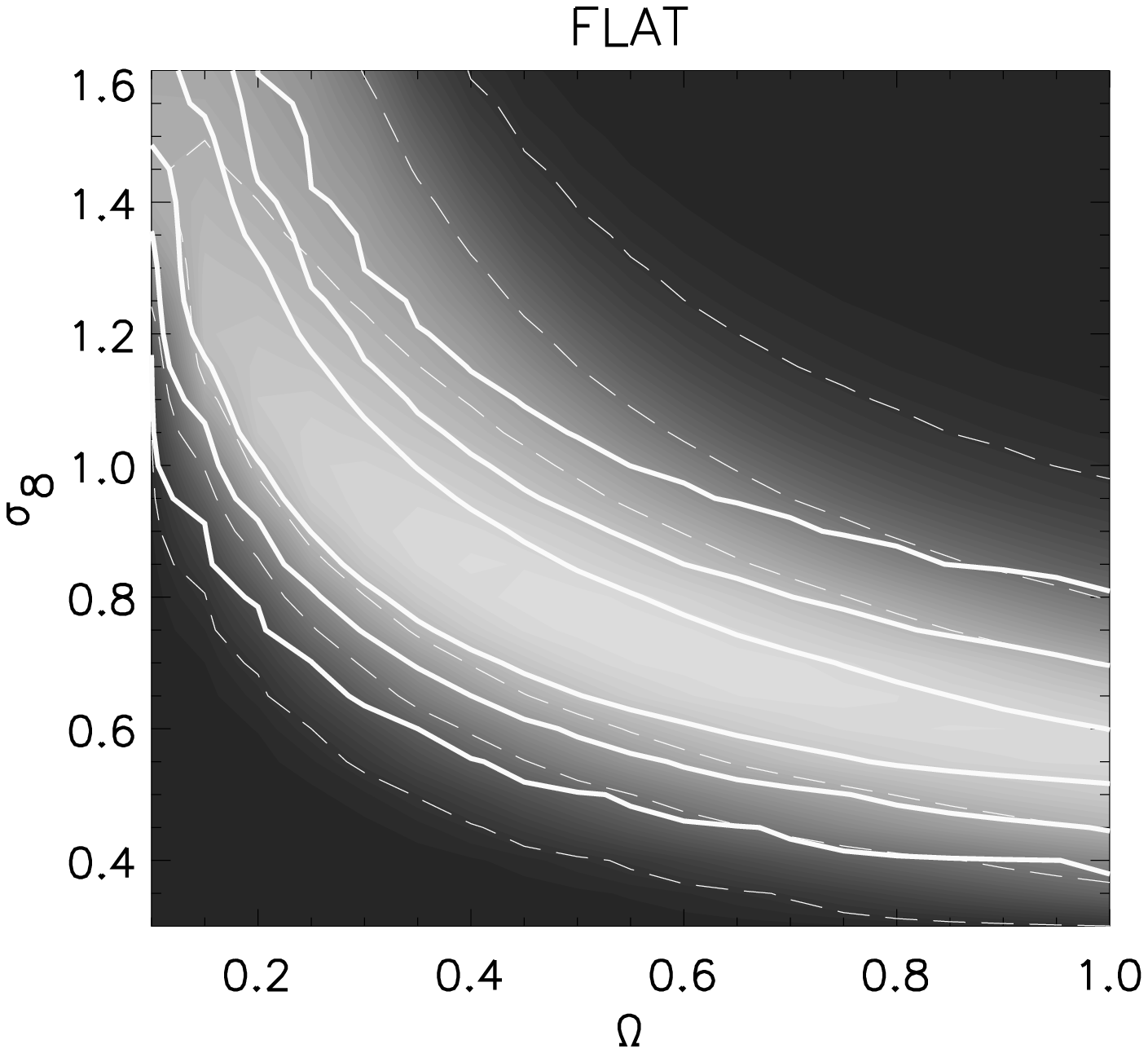,height=7cm}
\psfig{figure=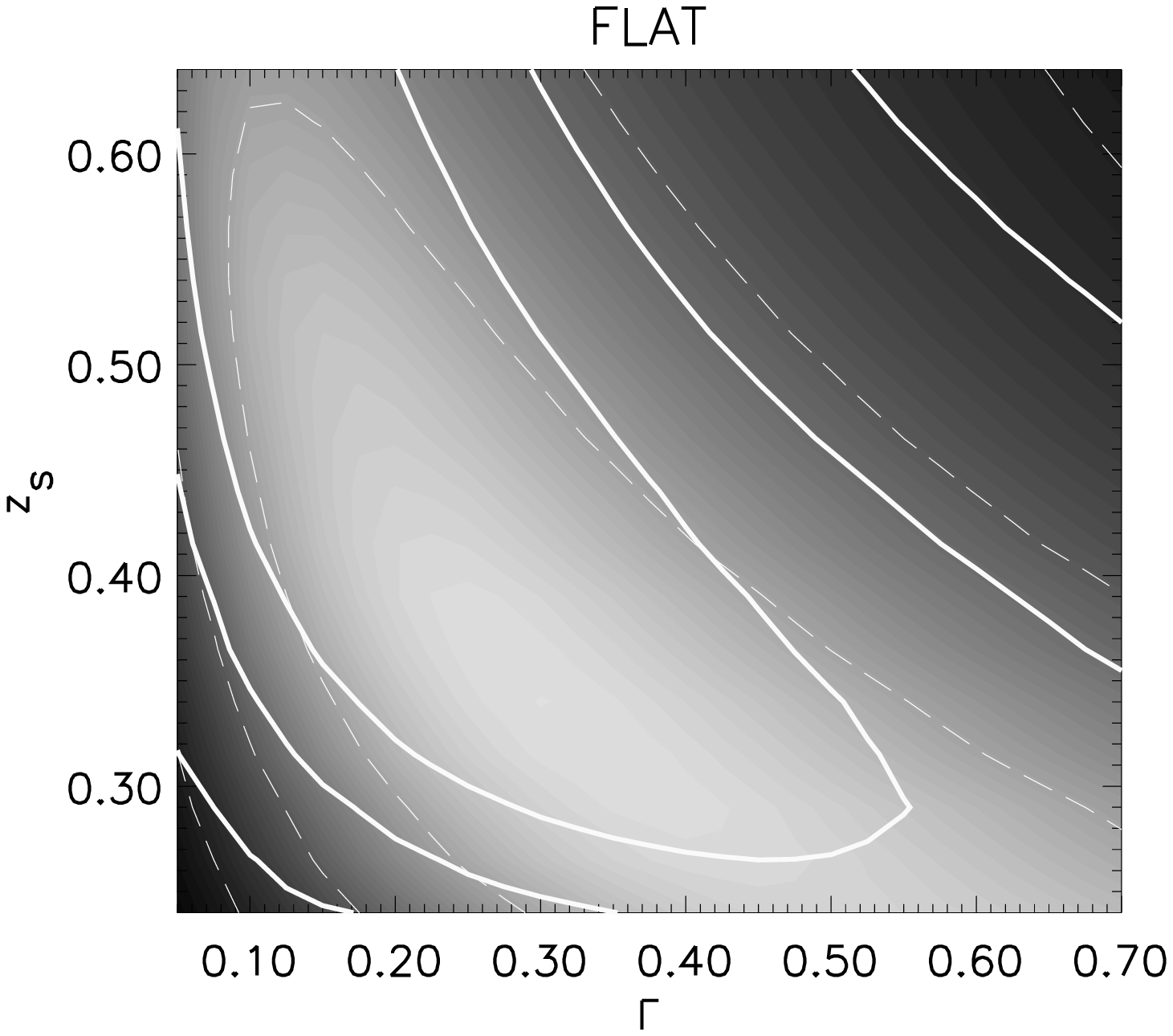,height=7cm}
}
\caption{\label{flat.ps} Left: constraints on
$\Omega$ and $\sigma_8$ for the flat cosmologies.
The confidence levels are $[68,95,99.9]$ from the brightest
to the darkest area. The galaxy sample has a magnitude cut $m_I>21$.
The gray area, and the dashed contours correspond to the 
contours computed with a full marginalization over the default
prior $\Gamma\in [0.05,0.7]$ and
$z_s\in [0.24,0.64]$. The thick solid line contours are
obtained from the prior $\Gamma\in [0.1,0.4]$ and $z_s\in [0.39,0.54]$
(which is a mean redshift $\bar z_s\in [0.8,1.1]$).
Right: constraints on
$\Gamma$ and $\Omega_M$ for the flat cosmologies. The contours have the same
statistical meaning as for the left panel, but here, the dashed lines
correspond to a marginalization over the default prior $\Omega \in [0.1,1]$
and $\sigma_8 \in [0.3, 1.6]$, and the thick lines for $\Omega \in [0.1,0.4]$
and $\sigma_8 \in [0.7, 1.3]$.
}
\end{figure*}

\begin{figure}
\centerline{
\psfig{figure=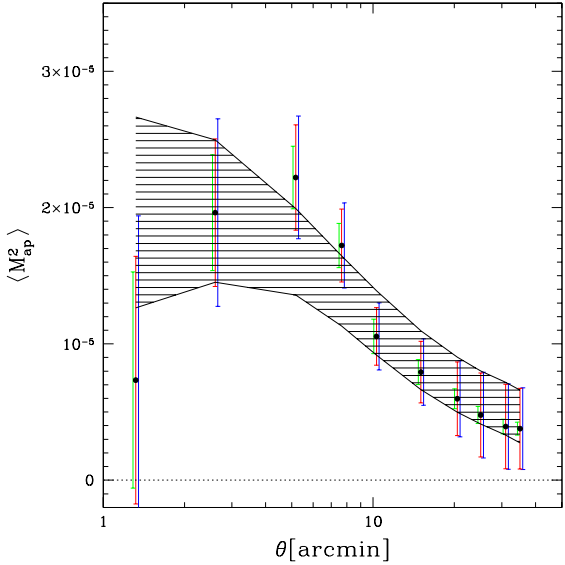,height=8cm}}
\caption{\label{data_models.ps} The aperture mass statistic
$\langle M_{\rm ap}^2\rangle$ measured on the data (see Figure
\ref{map_signal.ps}) compared to all the models included in the
$68\%$ contour (shaded area). For each measurement point, the
error bars from left to right are: statistical errors, statistical error
and residual bias, statistical errors and bias and cosmic variance.
}
\end{figure}

\begin{equation}
\sigma_8=\left(0.57\pm0.04\right) \Omega_M^{\left(0.24\mp 0.18\right)
\Omega_M-0.49},
\end{equation}
for the $68\%$ level and

\begin{equation}
\sigma_8=\left(0.58\pm0.13\right) \Omega_M^{\left(0.205\mp 0.025\right)
\Omega_M-0.48}
\end{equation}
for the $95\%$ contour. Constraints on the mass power spectrum can be
obtained from the right panel if one assume that photometric redshifts
provide the exact redshift distribution (which is given by $z_s=0.44$).
In that case we have $\Gamma \in [0.12,0.38]$.

Figure \ref{data_models.ps} shows the aperture mass measurements with
all the models inside the $68\%$ contours as the shaded area. The error
bars show the contribution the three errors as a function of scale. Each
set of errors shows 3 bars, which from left to right are: statistical noise,
bias added, cosmic variance added. We see that the statistical noise dominates
at small scale, while the systematic residuals dominate at larger scales, the
cosmic variance is never an important contribution.

\begin{figure}
\centerline{
\psfig{figure=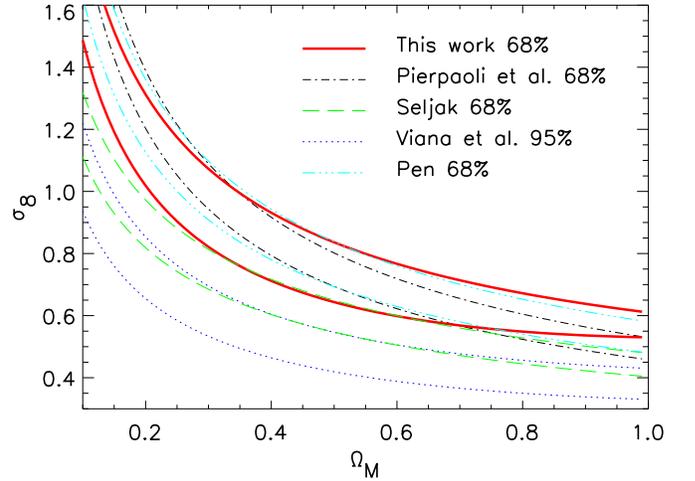,height=7cm}}
\caption{\label{sigma8omega.ps} The $\Omega-\sigma_8$ constraints for a
flat universe from
our work, compared to the cluster normalization constraints.
}
\end{figure}
\begin{figure}
\centerline{
\psfig{figure=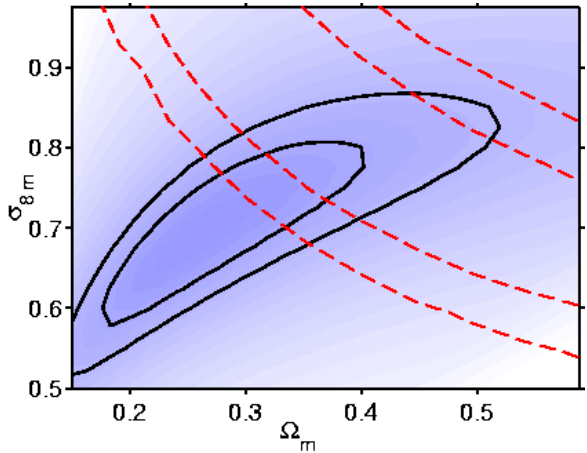,height=6cm}}
\caption{\label{cmblensing.ps} Solid line: the CMB alone constraints as
described in \cite{2001astro.ph.12162L} (figure borrowed from their
paper). The CMB priors are given by a concordance model, with
the Hubble constant $h=0.7$ with an r.m.s. of $0.07$, 
the baryon density $\Omega_b=0.02$, the primordial spectral index $n=1$,
and the reonization depth $\tau=0$.
Dashed lines our cosmic shear
constraints for the flat cosmologies (see Figure \ref{flat.ps}). In either case,
the contours show the $68\%$ and $95\%$ confidence levels.
}
\end{figure}
\begin{figure*}
\centerline{
\psfig{figure=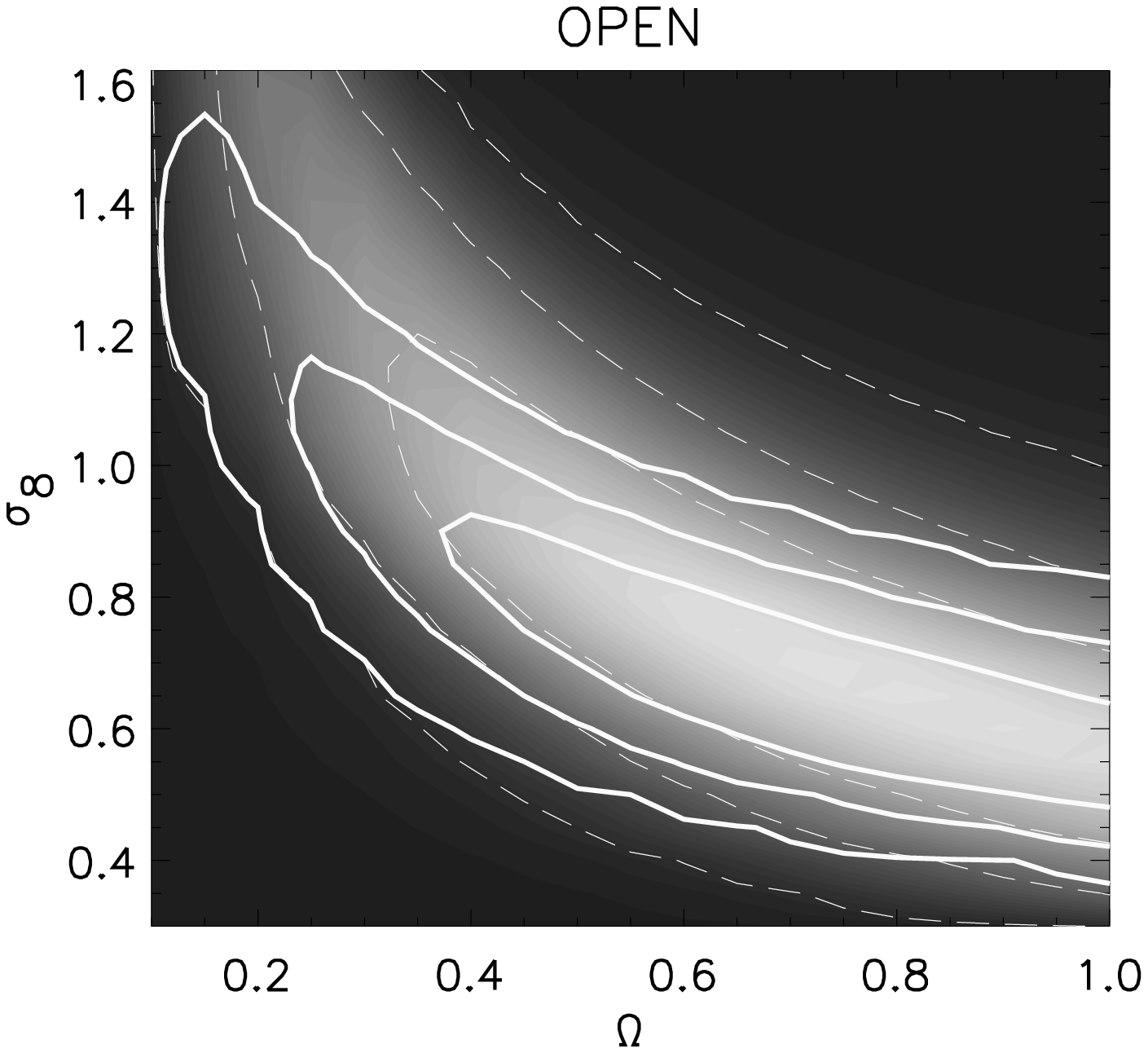,height=7cm}
\psfig{figure=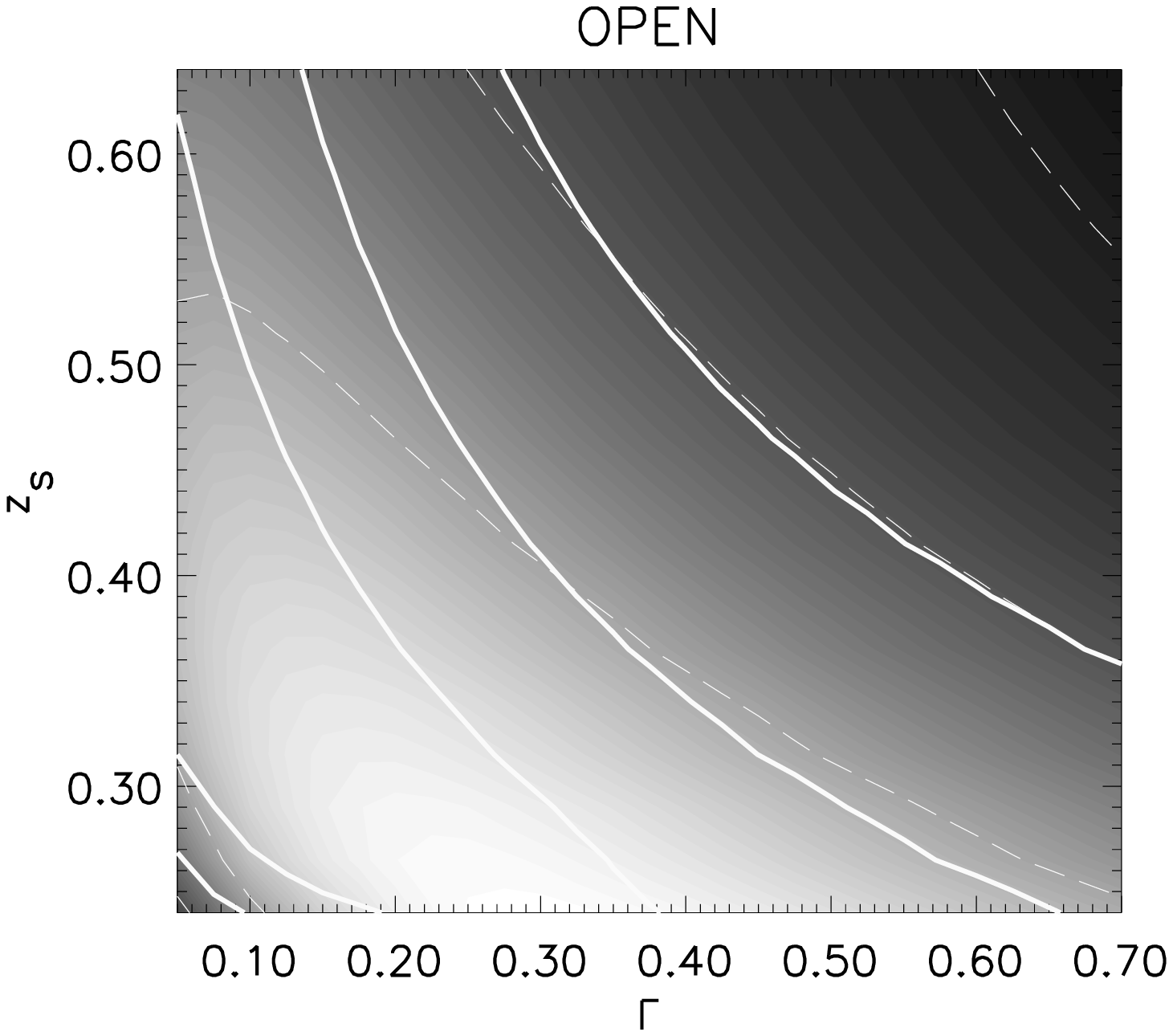,height=7cm}
}
\caption{\label{open.ps} Same as Figure \ref{flat.ps} for the open cosmologies.
}
\end{figure*}

It is interesting at this stage to compare our results with measurements
from other surveys. A comparison with Cosmic Microwave Background (CMB)
constraints reveals that weak lensing will be helpfull to break the degeneracy
between $\sigma_8$ and $\Omega_M$. Recently, \cite{2001astro.ph.12162L} have shown
CMB estimation of these two parameters, assuming that the Hubble constant
is a Gaussian variable centered at $h=0.7$ with an r.m.s. of $0.07$, and fixing
other parameters (primordial spectral index $n=1$, baryon density $\Omega_b=0.02$
and reionization depth $\tau=0$). Their results are shown as the solid lines
on Figure \ref{cmblensing.ps}. An overlay of our constraints on the same plot
(dashed lines) show that a combination of CMB and lensing would favorise low
density models ($\Omega_M \sim 0.3-0.4$) and rather low normalisation
($\sigma_8 \sim 0.7-0.8$). This plot reveals that the lensing constraints are
almost orthogonal to the CMB constraints. As explained in \cite{2001astro.ph.12162L},
a weaker prior on $h$ would extend the CMB contours and restore the degeneracy
between $\Omega_M$ and $\sigma_8$, making $\Omega_M=1$ a viable solution again. But
we see that lensing rules out such a solution because $\Omega_M \sim 1$ with
$\sigma_8 >0.8$ is excluded. Given that CMB alone predicts a flat Universe, the
inconsistency between CMB and lensing for $\Omega_M=1$ should be interpreted
as in favor of a non-zero cosmological
constant. The fact that CMB and lensing have
oposite constraints in the ($\sigma_8$,$\Omega_M$) parameter space make them
indeed very complementary. A complete analysis which take into account
the marginalization over the other parameters (baryon density,
$\tau$, etc...) is under way. However, we should note the agreement between
our results and the combined CMB+2dF contraints \citep[ Figure 5]{2001astro.ph.12162L}.

We should also compare our results more closely to the cluster
normalization constraints, since those two methods are expected
to probe a similar combination of $\sigma_8$ and $\Omega_M$.
Figure \ref{sigma8omega.ps} shows our
results and those obtained from cluster measurements. As it was claimed
before \citep{2001A&A...368..766M,
  2001A&A...374..757V,2001ApJ...552L..85R}, a joint estimate of
$\Omega_M$ and $\sigma_8$ from weak lensing is consistent with the
former cluster abundance estimates
\citep{1998ApJ...498...60P,2001MNRAS.325...77P}.  Recently, these
  estimates were revisited
 but the new results are 
puzzling (see Figure
\ref{sigma8omega.ps}): whereas \cite{2001astro.ph.11362S} is marginaly consistent 
with our constraints, on the other hand,
  \cite{2001astro.ph.11394V} is significantly lower.
\cite{1998ApJ...498...60P} performed direct hydrodynamic simulations to predict
the cluster X-ray temperature function for various cosmological
models.  This bypasses the difficult mass ladder, of converting
N-body or Press-Schechter mass functions into a temperature function,
and/or accounting for scatters in the relation, and the results 
  are in good agreement with the cosmic shear 
  constraints. However,  some effects
that may still not be accounted for in simulations include
non-gravitational feedback from galaxies, magnetic fields, thermal
conduction, which may all limit the intrinsic accuracy of cluster
normalizations.
We will not enter into the debate between the
cluster estimates here, but if the low normalization is confirmed, this
discrepancy might be an important finding: it might be an indication of
the inaccuracy of the non-linear predictions, as shown in Section 5.1.

The maximum likelihood analysis was also carried out for open
cosmologies.  The probability contours shown in Figure \ref{open.ps}
summarize the results which are indeed similar to the flat case.
However, low density ($\Omega_M < 0.2$), open universes, seem more
difficult to reconcile with the data than flat models.  This
contradiction between observations and low density open universe
results mostly from the small scale measurement of
$\langle M_{ap}^2\rangle$
(Figure \ref{data_models.ps}).  Indeed, low
density open universes predict too much power at small scale as
compared to what can be allowed from the amplitude of $\langle M_{ap}^2\rangle$
on scale of about one arc-minute.  This clear difference between open
and flat $\Lambda$CDM universes was already pointed out by
\cite{1998MNRAS.296..873S} (Fig. 3) but this is the first time that it
manifests on real data.

Finally, although uncertainties are still large enough to leave room
for a large sample of models, it is interesting to show how cosmic
shear data can be used jointly with several independent surveys
  (not only CMB).  Because weak
lensing analyses probe dark matter in a direct way, cosmic shear are
the best suited surveys to constrain $\sigma_8$.  It is then relevant
to only focus on this parameter, using values of other cosmological
parameters as they are derived from external data sets.  Assuming the
mean redshift of sources is $z_s=0.9$ and $h=0.7\pm 0.1$ \citep[ from
the HST Key Project]{Freedmanetal01}, a flat universe (from CMB data)
with a baryon fraction inferred from BBN and $\Gamma=0.2\pm0.05$
\citep[ from the SDSS redshift survey]{Szalayetal02}, we then have
$\Omega_M \approx 0.3\pm 0.1$.  In that case, the VIRMOS-DESCART cosmic
shear survey provide $\sigma_8=0.98 \pm 0.06$, in good agreement with
other independent methods. As compared to the join CMB-cosmic shear 
  alone discussed previously, the normalisation is higher. This is 
  mainly due to the low $\Omega_M$ (i.e. high $\sigma_8$) combined with
a strong prior on $\Gamma$ (inferred from the galaxy redshift
survey) and on the source redshift.

\section{Conclusion}

We explored a 4-dimensional parameter space using the most recent
cosmic shear data. We included all possible sources of error:
statistical noise, cosmic variance and residual systematics. We
obtained constraints on $\Omega_M$, the power spectrum slope $\Gamma$,
its normalization $\sigma_8$ and the redshift of the sources $z_s$. We
marginalized over $\Gamma$ and $z_s$. Both the marginalization, and the
inclusion of all the sources of error, make our results for
$(\Omega_M,\sigma_8)$ robust.
We pointed out the complementarity between cosmic shear and CMB
measurements for breaking the degeneracy among $\Omega_M$ and $\sigma_8$,
and the good agreement with CMB and CMB+2dF constraints.  
However, our results are only in marginal agreement with
the latest cluster abundance constraints, which give a lower
normalization $\sigma_8\sim 0.7$ for $\Omega_M\sim 0.3$.
If this discrepancy is confirmed in either
measurements, this could be interpreted as an indication that the
lensing non-linear prediction is not accurate enough given the
already small size of cosmic shear errors.
This interpretation is supported by ray-tracing simulations
in a $\tau$CDM model, and more generaly by a comparison of the VIRGO
simulations with the Peacock \& Dodds non-linear prescription.
It was claimed a $15\%$ accuracy in their original paper
\citep{1996MNRAS.280L..19P}, although it might be a bit more for some
cosmological models \citep{1998ApJ...499...20J}. This is clearly the
maximum uncertainty we can tolerate with today's lensing measurements,
and it will be insufficient for forthcoming surveys. This potential
problem suggests three paths for improvements:
\begin{enumerate}
\item Ray-tracing simulations should be used more intensively to test
  non-linear schemes. So far, simple ray-tracing have been performed,
  by putting all the sources at a single redshift ($z_s=1$), although
  in principle we should expect a redshift dependence of the non-linear
  predictions failure (because different physical scales are probes for
  a varying redshift, for a fixed angular scale).
\item Progress on the theory side should be done. There might be some
  hope by reviving the halo models which give predictions close, but
  not identical, to Peacock \& Dodds (like the peakpatch approach,
  Bond, priv. comm.).
\item Ultimately, cosmic shear observations will lead to a measurement
  of the 3D mass power spectrum in non-parametric way, and therefore
  solve all the problems associated with non-linear modeling.  This is
  possible only if the cosmological parameters are determined by other
  means. For instance, the linear mass power spectrum and cosmological
  parameters measurement at large scales using combined (or not)
  lensing data with cosmic microwave background, X-rays, could be
  obtained, and used to deconvolve the non-linear power spectrum.  This
  means that we will be able to deconvolve the projected mass power
  spectrum measured from cosmic shear observations and recover the true
  3D power spectrum. This is a work in progress, in which we are trying
  to recover the galaxy-galaxy and galaxy-mass correlations as well,
  using tomography techniques \citep{1999ApJ...522L..21H}.
\end{enumerate}

{ \acknowledgements We thank Dmitri Pogosyan and Carlo Contaldi for
  useful discussions maximum likelihood techniques and Peter Schneider
  and Henk Hoekstra for discussions and comments on the manuscript.
  Discussions with Roman Scoccimarro on the Peacock \& Dodds
  prescription were also very useful. We thank the VIRMOS and Terapix
  teams who got and processed the VIRMOS-DESCART data.  This work was
  supported by the TMR Network ``Gravitational Lensing: New Constraints
  on Cosmology and the Distribution of Dark Matter'' of the EC under
  contract No. ERBFMRX-CT97-0172. YM thanks CITA for hospitality.  }


\end{document}